\documentclass[12pt]{article}
\newcommand{\Bbb}[1]{\mathbf{#1}}
\usepackage[dvips]{graphicx}
\makeatletter
 
 \@addtoreset{equation}{section}
\makeatother 
\newtheorem{theorem}{Theorem}[section]
\newtheorem{lemma}[theorem]{Lemma}
\newcommand{\qed}{\rule{6pt}{6pt}}
\newcommand{\calP}{\mathcal{P}}
\newcommand{\up}{\uparrow}
\newcommand{\dn}{\downarrow}
\newcommand{\bs}{\backslash}
\newcommand{\bfs}{\mbox{\boldmath $\sigma$}}
\newcommand{\sgn}{{\mathbf{sgn}}}
\newcommand{\La}{\Lambda}
\newcommand{\rme}{\mathrm{e}}
\newcommand{\rmi}{\mathrm{i}}
\newcommand{\Ne}{{N_\mathrm{e}}}

\newcommand{\PhiGNe}{\Phi_{\mathrm{G},N_\mathrm{e}}}
\newcommand{\Tc}{T_\mathrm{c}}
\newcommand{\eqref}[1]{(\ref{#1})}
\begin{document}
\begin{center}
\textbf{\Large 
 A Model of Strongly Correlated Electrons 
 with Condensed Resonating-Valence-Bond
 Ground States 
}\bigskip\\
Akinori Tanaka%
\footnote{
 t-akitap@mbox.nc.kyushu-u.ac.jp
}\bigskip\\
\textit{Department of Applied Quantum Physics, Kyushu University,
 Fukuoka 812-8581, Japan}\bigskip\\
(September 12, 2003)
\end{center}
\vspace*{4cm}
\begin{abstract}
We propose a new exactly solvable model 
of strongly correlated electrons.
The model is based on a $d$-$p$ model 
of the CuO$_2$ plane with
infinitely large repulsive interactions 
on Cu-sites, and it contains 
additional correlated-hopping, pair-hopping and charge-charge 
interactions of electrons. 
For even numbers of electrons 
less than or equal to 2/3-filling,
we construct 
the exact ground states of the model,
all of which have the same energy 
and each of which is the unique ground state for 
a fixed electron number.  
It is shown that 
these ground states are the resonating-valence-bond states
which are also regarded as condensed states in which 
all electrons are in a single two-electron state. 
We also show that the ground states exhibit off-diagonal
long-range order.
\end{abstract}
\newpage
\section{Introduction}
Considerable attention has been paid to strongly correlated-electron
systems in the study of high-$\Tc$ superconductivity 
since Anderson~\cite{Anderson87}
proposed that its origin may be attributed to 
magnetic interactions induced by 
strong Coulomb repulsion.
In Anderson's scenario, 
a Mott insulating phase of an undoped cuprate is assumed to 
be the resonating-valence-bond (RVB) state, composed of 
electron spin-singlet pairs, 
and these singlet pairs become
charged superconducting pairs when the Mott insulator 
is doped sufficiently.  
Although 
many attempts have been made on lattice models of 
electrons with interactions,
such as the Hubbard model
and the $t$-$J$ model, 
there is no definite answer to the question of 
whether this scenario is realized 
in these concrete microscopic models.
These models are usually difficult
to analyze theoretically, and 
exact results are limited
except for one dimensional case 
where Bethe ansatz is available.

One of the exact results in two or more dimensions 
is obtained by Brandt and Giesekus~\cite{BG92},
who constructed the exact ground state of 
the Hubbard model with infinitely large repulsive on-site 
interactions on a $d$-dimensional decorated hypercubic lattice.
Tasaki~\cite{Tasaki93} generalized 
the Brandt-Giesekus model to a class of Hubbard
models on certain cell structures.
He pointed out that the exact ground states in this class of models 
are regarded as RVB states.
He also obtained the singlet pair correlation function on the tree
lattice and discussed a possibility 
of long-range order associated with
the singlet pairs and superconductivity.
Bares and Lee~\cite{BL94} and 
Yamanaka \textit{et al.}~\cite{Yamanaka96} treated
one dimensional versions of the models and obtained various
correlation functions, including the singlet pair correlation
function. 
They found that all the correlation functions decay exponentially
with distance 
and concluded that the system is in an insulating phase.
There are no conclusive results about the occurrence (or absence) of
superconductivity in the models in two or more dimensions.

For extended Hubbard models including nearest-neighbor
interactions in addition to the on-site interactions,
there are some exact results relevant to superconductivity. 
These results are based on the so-called $\eta$-pairing mechanism
originally proposed by Yang~\cite{Yang89}.
By using the $\eta$-pairing mechanism Yang 
constructed eigenstates of the Hubbard model on the hypercubic
lattice 
and proved that these states, in which all electrons form the $\eta$-pairs, 
have off-diagonal long-range order 
and thus are superconducting. 
Unfortunately, these eigenstates were shown not 
to be the ground states.  
But, Essler \textit{et al.}~\cite{Essler93} showed that 
the $\eta$-pairing states become the exact
ground states of an extended Hubbard model with attractive 
on-site interactions.
They also showed that
the model with moderately repulsive on-site interactions
has ground states in which  part of electrons 
forms the $\eta$-pairs.  
Possibilities of superconductivity 
due to the $\eta$-pairing mechanism
were discussed in other related models, 
in particular, in one dimensional systems.
See, for example, Refs.~\cite{BKS95}, \cite{KO02} and \cite{GG03} 
and references therein for more information. 

Another rigorous result is a theorem on the Hubbard model
with attractive on-site interactions on bipartite lattices 
due to S.-Q. Shen and Z.-M. Qiu~\cite{SQ93}. 
They proved that these models exhibit off-diagonal long-range
order in the ground states.
See also Ref.~\cite{Shen96} for related results.  

At the present time, 
to draw an exact result about superconductivity 
in the Hubbard model with on-site repulsion
or the $t$-$J$ model seems to be 
a formidably difficult task, 
but we have a chance to obtain one in extended Hubbard models.
Although these exactly solvable 
models have somewhat artificial aspects,
they will give us some insight into mechanisms 
for the phenomenon.   
In this paper,
motivated by this viewpoint,
we propose a new exactly solvable model of
strongly correlated electrons.
It is based on a two dimensional $d$-$p$ model,
a tight binding model of electrons with on-site interactions
on the decorated square lattice corresponding 
to the CuO$_2$ plane. 
The present new model 
has the infinitely large repulsive interactions on Cu-sites
and furthermore contains
correlated-hopping, pair-hopping and 
attractive charge-charge 
interactions of electrons.
For even numbers of electrons
we obtain the exact ground states of the model.
A mechanism used to construct the ground states
is similar to $\eta$-pairing mechanism, 
but an electron-pair used here
is different {from} $\eta$-pair.
Our electron-pair consists of singlet pairs of electrons
on the same O-site, on the nearest-neighbor pair of the Cu-sites, and 
on the nearest-neighbor pair of the Cu- and O-sites.
The constructed exact ground states are expressed as 
a linear combination of products of these singlet pairs 
and thus are regarded as
RVB states. 
We also show that these ground states exhibit off-diagonal
long-range order, which allow us to construct 
a  ground state
with explicit electron-number symmetry breaking.
 
The recent experimental results
of angle-resolved photoemission spectroscopy~\cite{Lanzara01}
suggested that electron-phonon coupling is
important in high-$T_\mathrm{c}$ superconductors and
thus a microscopic theory should include this effect.
Although this fact does not immediately justify 
an addition of artificial interactions 
to a model Hamiltonian,
we think that it is important to know
what kind of electron-electron interaction,
whichever of Coulomb repulsion or 
electron-phonon coupling (or others) induces this,
may stabilize a superconducting state
in a theoretical point of view.
We hope that our results 
will be useful in a future study.

In the next section 
we define the model and state the main result
as a theorem.
In Section~\ref{s:Remarks}, we give some remarks, 
including the RVB representation and off-diagonal 
long-range order. 
In Section~\ref{s:Proof} we prove the theorem.    
       
\section{Definition of the model and the main result}
We start by describing lattice~$\La$ on which our model 
will be defined.
Let $L$ be an arbitrary positive integer and 
let $D$ be a set of sites
\begin{equation}
 D = \{x=(x^{(1)},x^{(2)})~|~\mbox{$x^{(l)}\in\Bbb{Z}$, 
$1\le x^{(l)}\le L$ for $l=1,2$}\}
\end{equation}
with periodic boundary conditions. 
Let
$P$ be a set of sites located at the mid-points of nearest neighbor
sites in $D$,
\begin{equation}
 P = \{u = x+\delta_{l}/2~|~\mbox{$x\in D$, $l=1,2$}\},
\end{equation}
where $\delta_{l}$ is the unit vector along the $l$-axis. 
Then we define $\La$ as $\La=D\cup P$, which mimics the CuO$_2$ plane.
For later use, we also define
\begin{equation}
 P_x=\{u~|~u\in P,|u-x|=1/2 \} 
\end{equation}
for $x\in D$
and
\begin{equation}
 D_u=\{x~|~x\in D,|x-u|=1/2 \}
\end{equation}
for $u\in P$.

Let $c_{i,\sigma}$ and $c_{i,\sigma}^\dagger$ be the annihilation and
the creation operators, respectively,  
for an electron with spin $\sigma$ 
at site $i\in\La$.
These operators satisfy the usual anticommutation relations
\begin{equation}
 \{c_{i,\sigma},c_{j,\tau}\} 
  = \{c_{i,\sigma}^\dagger,c_{j,\tau}^\dagger\} = 0
\end{equation}
and
\begin{equation}
 \{c_{i,\sigma},c_{j,\tau}^\dagger\} = \delta_{ij}\delta_{\sigma\tau} 
\end{equation}
for any $i,j\in \La$ and $\sigma,\tau = \up,\dn$.
The number operator $n_{i,\sigma}$ is defined as
$n_{i,\sigma}=c_{i,\sigma}^\dagger c_{i,\sigma}$.
We denote by $\Ne$ the number of electrons in $\La$ and by $\Phi_0$ 
a state with no electrons.

We assume that repulsive interactions between electrons on $D$-sites 
are infinitely large and that each of these sites is 
at most singly occupied.
So we consider the Hilbert space $\mathbf{H}_\Ne$ spanned by the
linearly independent states of the form
\begin{equation}
\label{eq:basis}
 \Psi(A,\bfs_A;B_\up,B_\dn)
=\left(\prod_{x\in A}c_{x,\sigma_x}^\dagger\right)
 \left(\prod_{u\in B_\up}c_{u,\up}^\dagger\right)
 \left(\prod_{u\in B_\dn}c_{u,\dn}^\dagger\right)\Phi_0
\end{equation}
with arbitrary subsets $A\subset D$, $B_\up,B_\dn\subset P$ such that
$|A|+|B_\up|+|B_\dn|=\Ne$.
Here $\bfs_A$ is a shorthand 
for a spin configuration $(\sigma_x)_{x\in A}$. 

To define the Hamiltonian, we introduce the following new fermion
operators:
\begin{eqnarray}
 a_{u,\sigma} =
  c_{u,\sigma} + \alpha\sum_{x\in D_u} c_{x,\sigma} 
& \mbox{for $u\in P$},\\
 b_{x,\sigma} = 
  c_{x,\sigma} + \beta\sum_{u\in P_x} c_{u,\sigma} 
& \mbox{for $x\in D$},\\
 d_{x,\sigma} = 
  \sum_{u\in P_x}\rme^{-2 \rmi Q\cdot u} a_{u,\sigma}\, 
& \mbox{for $x\in D$},
\end{eqnarray}
where $\alpha$ and $\beta$ are real parameters 
with $\alpha\ne-\beta$, 
and $Q=(0,\pi)$.
We also introduce operators 
$n_{u,\sigma}^{a}=a_{u,\sigma}^\dagger a_{u,\sigma}$,
$n_{x,\sigma}^{b}=b_{x,\sigma}^\dagger b_{x,\sigma}$ and
$n_{x,\sigma}^{d}=d_{x,\sigma}^\dagger d_{x,\sigma}$.  
Then, by using these operators, we define
\begin{equation}
 H_0 = \calP_D \left(t\sum_{x\in D}\sum_{\sigma=\up,\dn} 
		n_{x,\sigma}^b
		+s \sum_{u\in P}\sum_{\sigma=\up,\dn}
		n_{u,\sigma}
	       \right)\calP_{D}, 
\end{equation}
\begin{equation}
 H_{\mathrm{int},1} 
  = \calP_D \left(-V_1\sum_{x\in D}\sum_{\sigma=\up,\dn} 
	     n_{x,-\sigma}^d n_{x,\sigma}^b
	     -W_1 \sum_{x\in D}\sum_{l=1,2}\sum_{\sigma=\up,\dn}
	     (n_{x+\delta_{l},\sigma}
	     +n_{x-\delta_{l},\sigma})n_{x,\sigma}^b
	    \right)\calP_{D}, 
\end{equation} 
and
\begin{equation}
 H_{\mathrm{int},2} 
  = \calP_D \left(-V_2\sum_{u\in P}\sum_{\sigma=\up,\dn} 
	     n_{u,-\sigma}^a n_{u,\sigma}
	     -W_2 \sum_{u\in P}\sum_{x\in D_u}
	     \sum_{\sigma=\up,\dn}n_{x,\sigma}n_{u,\sigma}
	    \right)\calP_{D}, 
\end{equation}
where $t,s,V_1,W_1,V_2$ and $W_2$ are real parameters, and
$-\sigma$ denotes the spin opposite to $\sigma$. 
The projection operator $\calP_{D}$ 
which eliminates states with doubly
occupied $D$-sites is defined by 
\begin{equation}
 \calP_{D} = \prod_{x\in D} \calP_x
\end{equation}
with 
\begin{equation}
 \calP_x=(1-n_{x,\up}n_{x,\dn}).
\end{equation}

\begin{figure}
 \begin{center}
 \includegraphics[scale=.8]{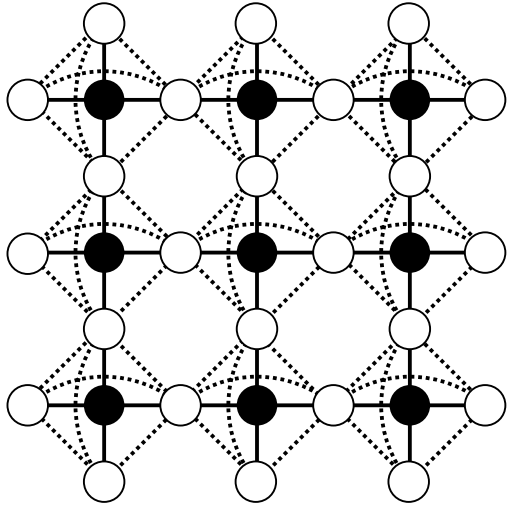}
 \end{center}
 \caption{
 The lattice structure and the hopping matrix elements of $H_0$.
 The filled circles and the open circles 
 correspond to $D$-sites and $P$-sites, 
 respectively. 
 The on-site potential of $D$-sites is $t$ and that of $P$-sites is
 $2\beta^2t +s$. 
 The solid lines and 
 the dashed lines represent the hopping matrix elements, $\beta t$   
 and $\beta^2 t$, respectively. 
 }
 \label{fig:lattice}
\end{figure}

The Hamiltonian $H_0$ is rewritten as
\begin{equation} 
 H_0=\calP_D\left(
  \sum_{i,j\in\La}t_{ij}c_{i,\sigma}^\dagger c_{j,\sigma}
  \right)\calP_D
\end{equation}
with
\begin{equation}
t_{ij}=\left\{
 \begin{array}{ll}
  t & \mbox{if $i=j\in D$;}\\
  2\beta^2t + s & \mbox{if $i=j\in P$;}\\
  \beta t & \mbox{if $|i-j|=\frac{1}{2}$;}\\
  \beta^2 t & \mbox{if $i\ne j$, $i,j\in P_x$ for some $x\in D$;}\\
  0 & \mbox{otherwise,}
 \end{array}
\right.
\end{equation}
(see Fig.~\ref{fig:lattice}) and 
it describes quantum mechanical motion of electrons feeling
infinitely large
repulsive interactions at $D$-sites.
The Hamiltonians $H_{\mathrm{int},1}$ and 
$H_{\mathrm{int},2}$
correspond to
correlated-hopping, pair-hopping, charge-charge,   
and spin-spin interactions.
We can observe this  
by rewriting, for example,
$H_{\mathrm{int},2}$ as 
\begin{eqnarray}
H_{\mathrm{int},2} 
&=&
 \calP_D 
 \Bigg( \sum_{u\in P}
 \Big(
 -\alpha^2 V_2
 \sum_{x,y\in D_u;x\ne y}\sum_{\sigma=\up,\dn}
 c_{x,-\sigma}^\dagger c_{y,-\sigma} n_{u,\sigma}
\nonumber\\
&&\hspace*{3em}
 -\alpha V_2 \sum_{x\in D_u}\sum_{\sigma=\up,\dn}
 (c_{x,-\sigma}^\dagger c_{u,-\sigma}
 +c_{u,-\sigma}^\dagger c_{x,-\sigma}) 
 n_{u,\sigma}
 -2V_2 n_{u,\up}n_{u,\dn}
\nonumber\\ 
&&\hspace*{3em}
 -\frac{1}{2}(\alpha^2V_2+W_2)
 \sum_{x\in D_u}n_{x}n_{u}
 +2(\alpha^2V_2-W_2)\sum_{x\in D_u}S_{x}^{(3)}S_{u}^{(3)}
 \Big)
 \Bigg)\calP_D,
\end{eqnarray}
where $n_i=n_{i,\up}+n_{i\dn}$ and 
$S_{i}^{(3)}=(n_{i,\up}-n_{i\dn})/2$.
The Hamiltonian $H_{\mathrm{int},1}$
can be rewritten as well, 
but that has a somewhat complicated form.
We note that the interactions $H_{\mathrm{int},1}$ 
and $H_{\mathrm{int},2}$
conserve the electron number 
and the eigenvalue of $\sum_{i\in\La}S_{i}^{(3)}$,
but they
do not possess the spin SU(2) symmetry.

In this paper we consider the Hamiltonian
\begin{equation}
 H=H_0 + H_{\mathrm{int},1} + H_{\mathrm{int},2} 
\end{equation} 
with the parameters satisfying $t=4(1+\alpha^2)V_1$,
$W_1=\alpha^2V_1$
$s=(1+2\alpha^2)V_2$, $W_2 = \alpha^2 V_2$ and $V_1,V_2>0$.
In this case, the direct spin-spin interaction terms
vanish and the charge-charge interactions are attractive. 

In our construction of exact ground states, it is crucial
to rewrite $H$ as
\begin{equation}
H=\sum_{x\in D} H_x + \sum_{u\in P} H_u 
\end{equation}
with 
\begin{equation}
 H_x = \calP_D\left(
	       V_1\sum_{\sigma=\up,\dn} 
	       b_{x,\sigma}^\dagger d_{x,-\sigma}
	       \calP_D
	       d_{x,-\sigma}^\dagger b_{x,\sigma} 
	      \right)\calP_D
\end{equation}
for $x\in D$ and 
\begin{equation}
 H_u = \calP_D\left(
	       V_2\sum_{\sigma=\up,\dn} 
	       c_{u,\sigma}^\dagger a_{u,-\sigma}
	       \calP_D
	       a_{u,-\sigma}^\dagger c_{u,\sigma} 
	      \right)\calP_D
\end{equation}
for $u\in P$.
Here we used the operator identities~\cite{BG92,Tasaki94}
\begin{equation}
 \label{eq:identity1}
 \calP_D c_{i,\sigma} \calP_D c_{j,\sigma}^\dagger \calP_D
  =\left\{
    \begin{array}{ll}
     -\calP_D c_{j,\sigma}^\dagger c_{i,\sigma} \calP_D & 
      \mbox{if $i\ne j$}; \\ 
     \calP_D (1-n_{i,\up}-n_{i,\dn}) \calP_D & 
      \mbox{if $i=j\in D$}; \\ 
     \calP_D (1-n_{i,\sigma}) \calP_D & \mbox{if $i=j\in P$},
    \end{array}
\right.
\end{equation}
and
\begin{eqnarray}
 \label{eq:identity2}
  \calP_{x} c_{i,\sigma} \calP_{x} 
  = c_{i,\sigma} \calP_{x}, \\
 \label{eq:identity3}
  \calP_{x} c_{i,\sigma}^\dagger \calP_{x} 
  = \calP_{x} c_{i,\sigma}^\dagger.
\end{eqnarray}
Operator identities \eqref{eq:identity2} and \eqref{eq:identity3}
follow {from} straightforward calculations, 
and \eqref{eq:identity1} is derived by using 
$n_{x,\up}n_{x,\dn}\calP_{x}=0$ and 
\begin{eqnarray}
\label{eq:identity4}
 c_{i,\sigma}\calP_{x} 
  =\left\{ 
    \begin{array}{ll} 
     (1-n_{x,-\sigma})c_{x,\sigma}
      =\calP_{x}(1-n_{x,-\sigma})c_{x,\sigma} &  
      \mbox{if $i=x$};\\
     \calP_{x}c_{i,\sigma} & \mbox{otherwise},
    \end{array}\right. \\
 \label{eq:identity5}
  \calP_{x}c_{i,\sigma}^\dagger 
  =\left\{ 
    \begin{array}{ll} 
        c_{x,\sigma}^\dagger(1-n_{x,-\sigma})
      = c_{x,\sigma}^\dagger (1-n_{x,-\sigma})\calP_{x} &  
      \mbox{if $i=x$};\\
     c_{i,\sigma}^\dagger\calP_{x} & \mbox{otherwise},
    \end{array}\right. 
\end{eqnarray}
which also follow {from} simple calculations.
We note that $H_i$ with $i\in\La$ are positive semidefinite
operators
and thus a zero-energy state for all 
$H_i$ (if it exists) is a ground state.   

Let us define
\begin{equation}
 \zeta^\dagger = 
  \sum_{u\in P}\rme^{2\rmi Q\cdot u}
  a_{u,\up}^\dagger a_{u,\dn}^\dagger,
\end{equation}
which corresponds to a singlet two-electron state.
The main result in this paper is the following theorem:
\begin{theorem}
\label{theorem}
Fix the electron number $\Ne$ and suppose that $\Ne \le 2|P|$. 
When $\Ne$ is even, the ground state $\PhiGNe$ of $H$
is unique and is given by
\begin{equation}
\label{eq:ground state}
 \PhiGNe = \calP_D\left(\zeta^\dagger \right)^{\frac{\Ne}{2}}\Phi_0
       = \left(\calP_D\zeta^\dagger \right)^{\frac{\Ne}{2}}\Phi_0,
\end{equation}
which satisfies $H\PhiGNe=0$.
For odd $\Ne$, the ground state energy is positive. 
\end{theorem}

It is noted that 
the ground state energy for odd $\Ne$ 
may converge to zero as $L\to\infty$.
Whether this is the case or not should be clarified
in a future study.  
\section{Some Remarks}
\label{s:Remarks}
It is easy to verify the relation
\begin{equation}
\zeta^\dagger
= 
	\sum_{x\in D}\sum_{l=1,2}\rme^{\rmi Q\cdot\delta_l}
	\left(
	 \alpha^2 f_{x,x+\delta_l}^\dagger 
	 +\alpha   f_{x,x+\delta_l/2}^\dagger
	 +\alpha   f_{x,x-\delta_l/2}^\dagger
	 \right)
	+ 
	\sum_{u\in P} \rme^{2 \rmi Q\cdot u} 
             c_{u,\up}^\dagger c_{u,\dn}^\dagger,
\end{equation}
where 
$f_{i,j}^{\dagger}=c_{i,\up}^\dagger c_{j,\dn}^\dagger 
     + c_{j,\up}^\dagger c_{i,\dn}^\dagger$
is the creation operator for 
the valence-bond (singlet pair) 
on sites $i$ and $j$.
Furthermore the projection operator $\calP_D$ eliminates 
the terms of the form 
\begin{equation}
 f_{x,i}^\dagger f_{x,j}^\dagger\cdots\Phi_0
  = - c_{x,\up}^\dagger c_{x,\dn}^\dagger f_{i,j}^\dagger\cdots\Phi_0,
\end{equation}
when we express $\PhiGNe$ 
by using $f_{i,j}^\dagger$ and 
$c_{u,\up}^\dagger c_{u,\dn}^\dagger$. 
Thus the ground state $\PhiGNe$ has the form of the RVB state
which is a linear combination of products of on-site singlet pairs 
$c_{u,\up}^\dagger c_{u,\dn}^\dagger$ in $P$
and
the valence-bonds $f_{x,x+\delta_l}^\dagger$ 
and $f_{x,x\pm\delta_l/2}^\dagger$.
It is noted that the weight for the singlet pairs
$f_{x,x+\delta_l}^\dagger$ 
on the nearest-neighbor $D$-sites becomes relatively large
for $1/\alpha \ll 1$.

The term $-V_1 n_{x,-\sigma}^d n_{x,\sigma}^b$ 
in interaction Hamiltonian
$H_{\mathrm{int},1}$ is rewritten as
\begin{equation}
 -V_1 n_{x,-\sigma}^d n_{x,\sigma}^b
  = -4(1+\alpha^2)(1+4\beta^2)V_1 
  + 4(1+\alpha^2)V_1 b_{x,\sigma}b_{x,\sigma}^\dagger
  + V_1 d_{x,-\sigma}b_{x,\sigma}^\dagger
        b_{x,\sigma} d_{x,-\sigma}^\dagger.
\end{equation}
{From} this expression,
we find that $-V_1 n_{x,-\sigma}^d n_{x,\sigma}^b$ is bounded
below by $-4(1+\alpha^2)(1+4\beta^2)V_1$ and furthermore that 
it is attained by states of the form 
$b_{x,\sigma}^\dagger d_{x,-\sigma}^\dagger\cdots\Phi_0$.  
Therefore this interaction term is describing 
an attractive and magnetic interaction between localized
single-electron states corresponding to fermion operators
$b_{x,\sigma}$ and $d_{x,\sigma}$.
The other terms can be rewritten into similar forms
and interpreted as attractive and magnetic interactions as well.
Our model says that attractive magnetic interactions between 
localized states together with 
the strong on-site repulsion
can stabilize the condensed RVB
states consisting of local singlet pairs.

The explicit expression of the ground states makes 
it possible to evaluate 
order parameters. 
We first note that, in the case of $\alpha=0$, $\PhiGNe$ is reduced 
to the $\eta$-pairing state on the $P$-sites, 
and therefore the ground states have 
off-diagonal long-range order.

For $\alpha\ne 0$ we can prove the existence of 
off-diagonal long-range order as follows.
Let us introduce order parameters
\begin{equation}
 \Delta_{1,\La} 
  = \calP_D
    \left(
     \frac{1}{|P|}\sum_{u\in P} 
     \rme^{-2\rmi Q\cdot u}c_{u,\dn}c_{u,\up}
    \right)
    \calP_D
\end{equation}
and
\begin{equation}
 \Delta_{2,\La} 
  = 
  \calP_D
  \left(
     \frac{1}{|P|}\sum_{u\in P} 
     \rme^{-2\rmi Q\cdot u}a_{u,\dn}a_{u,\up}
  \right)
  \calP_D
  =\frac{1}{|P|}\calP_D\zeta\calP_D.
\end{equation}
Noting the anticommutation relation 
\begin{equation}
\label{eq:anticommutation1}
  \{c_{u,\sigma},a_{u^\prime,\sigma}^\dagger\}
  =\delta_{uu^\prime}
\end{equation}
for $u,u^\prime\in P$, 
one finds the commutation relation
\begin{equation}
\label{eq:commu-delta1-2}
 [\Delta_{1,\La},\calP_D\zeta^\dagger\calP_D] 
  =
  \calP_D
  \left(
  \frac{1}{|P|}
  \sum_{u\in P}(1-
  a_{u,\up}^\dagger c_{u,\up} 
  - a_{u,\dn}^\dagger c_{u,\dn})
  \right)
  \calP_D
\end{equation}
and
\begin{equation}
  \calP_D\sum_{u\in P}(1-
   a_{u,\up}^\dagger c_{u,\up} 
   - a_{u,\dn}^\dagger c_{u,\dn} )\calP_D\PhiGNe
   = (|P|-\Ne)\PhiGNe.
\end{equation}
These two relations lead to
\begin{equation}
 \label{eq:eigenDelta1}
 \Delta_{1,\La}\PhiGNe 
  = \left(
     \frac{\Ne}{2}
     -
     \frac{1}{|P|}\frac{\Ne}{2}
     \left(
      \frac{\Ne}{2}-1
     \right)
    \right)\Phi_{\mathrm{G},\Ne-2}.
\end{equation}
Therefore we obtain
\begin{equation}
\label{eq:Delta12}
 \langle \Delta_{2,\La}^\dagger \Delta_{1,\La} \rangle_{\La,\Ne}
  =
  \frac{1}{|P|}
  \left(
   \frac{\Ne}{2}
   -\frac{1}{|P|}
   \frac{\Ne}{2}\left(\frac{\Ne}{2}-1\right)
  \right),
\end{equation}
where the expectation value $\langle \cdots \rangle_{\La,\Ne}$ is
defined by
\begin{equation}
 \langle \cdots \rangle_{\La,\Ne}
   =
  \frac{\left(\PhiGNe,\cdots\PhiGNe\right)}
  {\left(\PhiGNe,\PhiGNe\right)}.
\end{equation}
We write $\mu_{\La,\Ne}$ 
for the right-hand-side of \eqref{eq:Delta12}.
Then, by using the Schwarz inequality 
\begin{equation}
 |\langle \Delta_{2,\La}^\dagger \Delta_{1,\La} \rangle_{\La,\Ne}|^2
  \le
  \langle \Delta_{2,\La}^\dagger \Delta_{2,\La} \rangle_{\La,\Ne}
  \langle \Delta_{1,\La}^\dagger \Delta_{1,\La} \rangle_{\La,\Ne},
\end{equation}
and inequalities 
$\langle \Delta_{1,\La}^\dagger \Delta_{1,\La}\rangle_{\La,\Ne} 
\le 1$ 
and 
$\langle \Delta_{2,\La}^\dagger \Delta_{2,\La}\rangle_{\La,\Ne} 
\le (1+2\alpha)^2$,
we find
\begin{equation}
 \label{eq:inequalityDelta1}
 \frac{\mu_{\La,\Ne}^2}{(1+2\alpha^2)^2}\le 
   \langle \Delta_{1,\La}^\dagger \Delta_{1,\La}\rangle_{\La,\Ne} 
   \le 1 
\end{equation}
and
\begin{equation}
 \label{eq:inequalityDelta2}
 {\mu_{\La,\Ne}^2}\le 
   \langle \Delta_{2,\La}^\dagger \Delta_{2,\La}\rangle_{\La,\Ne} 
   \le (1+2\alpha^2)^2.
\end{equation}
Let $\{\Phi_{{\mathrm{G}},L}^{(\nu)}\}_{L=0}^\infty$ be a sequence of
ground states $\PhiGNe$ on the lattice with side-length $L$ such that  
the electron filling $\Ne/(2|\La|)=\Ne/(6L^2)$ 
converges to $\nu$ as $L\to\infty$,
and let $\{\langle \cdots \rangle_{L}^{(\nu)}\}_{L=0}^\infty$
be a sequence of expectation values with respect to 
$\Phi_{{\mathrm{G}},L}^{(\nu)}$. 
Then {from} inequalities \eqref{eq:inequalityDelta1} 
and \eqref{eq:inequalityDelta2}
we have 
\begin{equation}
 \liminf_{L\to\infty} \langle 
  \Delta_{1,\La}^\dagger\Delta_{1,\La}
  \rangle_{L}^{(\nu)}
  \ge \frac{\mu_\nu^2}{(1+2\alpha^2)^2}
\end{equation}
and
\begin{equation}
 \liminf_{L\to\infty} \langle 
 \Delta_{2,\La}^\dagger \Delta_{2,\La}
 \rangle_{L}^{(\nu)}
 \ge
 \mu_{\nu}^2
\end{equation}
where $\mu_\nu=\frac{3\nu}{2}\left(1-\frac{3\nu}{2}\right)$,
which imply the existence of off-diagonal long-range order
for $0<\nu<2/3$.

Now we can construct a ground state
with explicit electron-number symmetry breaking~\cite{KT94}.
Let us define
\begin{equation}
\label{eq:breaking-state}
\Phi_{\mathrm{G},\Ne}^\prime = \Phi_{\mathrm{G},\Ne}
              + 
	      \frac{\Delta_{2,\La}^\dagger}
	      {\sqrt{\langle \Delta_{2,\La} 
	      \Delta_{2,\La}^\dagger \rangle_{\La,\Ne}}}\PhiGNe,
\end{equation}
which is also a zero-energy state.
We note that the limit infimum of 
$\langle \Delta_{2,\La} \Delta_{2,\La}^\dagger \rangle_L^{(\nu)}$ is
bounded below by $\mu_\nu^2$, since 
\begin{equation}
 \langle \Delta_{1,\La} \Delta_{2,\La}^\dagger \rangle_{\La,\Ne} 
  = \langle \Delta_{2,\La}^\dagger \Delta_{1,\La} \rangle_{\La,\Ne} 
  + \frac{1}{|P|^2}(|P|-\Ne),
\end{equation}
which follows {from} \eqref{eq:commu-delta1-2}.
Since two states with different electron number are orthogonal,
we find that
\begin{equation}
 \langle \Delta_{2,\La} \rangle_{\La,\Ne}^\prime 
  = 
  \frac{\left(\Phi_{\mathrm{G},\Ne}^\prime,
	 ~\Delta_{2,\La} 
	 \Phi_{\mathrm{G},\Ne}^\prime\right)}
  {\left(\Phi_{\mathrm{G},\Ne}^\prime,
    ~\Phi_{\mathrm{G},\Ne}^\prime\right)}
  =
  \frac{1}{2}
  \sqrt{\langle \Delta_{2,\La} \Delta_{2,\La}^\dagger 
  \rangle_{\La,\Ne}}.
\end{equation}
Therefore, the limit infimum of
the sequence of $\langle \Delta_{2,\La} \rangle_{\La,\Ne}^\prime$
obtained by using $\{\Phi_{{\mathrm{G}},L}^{(\nu)}\}_{L=0}^\infty$
with \eqref{eq:breaking-state}    
is bounded below by $\mu_{\nu}/2$, 
which implies electron-number symmetry breaking.

We finally remark about extensions of the present model.
It is possible to construct similar models 
in one, three and more dimensions  
by the same method.
The model in three or more dimensions may 
exhibit a finite temperature phase transition.
It is also possible to construct models
whose ground states are written as 
$\left(\zeta^\dagger\right)^{\frac{\Ne}{2}}\Phi_0$. 
The details will appear elsewhere.
\section{Proof}
\label{s:Proof}
\textit{Proof of Theorem~\ref{theorem}.} 
We first prove that $\PhiGNe$ in \eqref{eq:ground state} is a
zero-energy state
for all $H_i$ and thus is a ground state.
Using the anticommutation 
relations~\eqref{eq:anticommutation1}
and
\begin{equation}
\label{eq:anticommutation2}
 \{c_{x,\sigma},a_{u,\sigma}^\dagger\}
  =\alpha\chi[u\in P_x]
\end{equation}
for $x\in D$,
where $\chi[\textrm{``event''}]$ takes 1 if ``event''
is true and takes 0 otherwise,
one finds the following two commutation relations:
\begin{equation}
\label{eq:commutation1}
[c_{u,\sigma},\zeta^\dagger] 
= \sigma \rme^{2\rmi Q\cdot u} a_{u,-\sigma}^\dagger
\end{equation}
for $u\in P$ and 
\begin{equation}
\label{eq:commutation2}
 [c_{x,\sigma},\zeta^\dagger]
  =\sigma\alpha
  \sum_{u\in P_x}
  \rme^{2\rmi Q\cdot u}a_{u,-\sigma}^\dagger
\end{equation}
for $x\in D$.
(In the right-hand-sides of \eqref{eq:commutation1} 
and \eqref{eq:commutation2}, the coefficients 
$\sigma=\up$ and $\dn$ are regarded as $+1$ and $-1$, respectively. 
We will use this convention in the following.) 
By using operator identities \eqref{eq:identity3},
\eqref{eq:identity4} 
and commutation relation \eqref{eq:commutation1}, 
we find 
\begin{eqnarray}
 (\calP_D a_{u,-\sigma}^\dagger c_{u,\sigma}) \calP_D \zeta^\dagger 
  &=& \calP_D a_{u,-\sigma}^\dagger c_{u,\sigma}\zeta^\dagger 
  \nonumber\\
  &=&  \calP_D a_{u,-\sigma}^\dagger
   (\zeta^\dagger c_{u,\sigma}
   +\sigma \rme^{2\rmi Q\cdot u} a_{u,-\sigma}^\dagger)\nonumber\\
  &=&  \calP_D \zeta^\dagger 
   (\calP_D a_{u,-\sigma}^\dagger c_{u,\sigma}).
\end{eqnarray}
To get the third line, we  used $(a_{u,-\sigma}^\dagger)^2=0$.
This implies that 
$(\calP_D a_{u,-\sigma}^\dagger c_{u,\sigma}) 
(\calP_D\zeta^\dagger)^{\frac{\Ne}{2}}\Phi_0 = 0$,
and thus $H_u \PhiGNe=0$ for all $u\in P$.
Similarly, by using \eqref{eq:identity3}, \eqref{eq:identity4} and
commutation relations \eqref{eq:commutation1}, 
\eqref{eq:commutation2} 
we have 
\begin{eqnarray}
 \left(\calP_D d_{x,-\sigma}^\dagger 
  \sum_{u\in P_x} c_{u,\sigma}\right) \calP_D \zeta^\dagger
&=& 
 \left(\calP_D d_{x,-\sigma}^\dagger 
  \sum_{u\in P_x} c_{u,\sigma}\right)\zeta^\dagger \nonumber\\
&=&
 \calP_D d_{x,-\sigma}^\dagger \zeta^\dagger 
 \sum_{u\in P_x} c_{u,\sigma} 
 +\sigma \calP_D d_{x,-\sigma}^\dagger \sum_{u\in P_x} 
 \rme^{2 \rmi Q\cdot u} a_{u,-\sigma}^\dagger
 \nonumber\\
 &=& 
\label{eq:PdxbxPhi1}
 \calP_D \zeta^\dagger
 \left(\calP_D d_{x,-\sigma}^\dagger 
  \sum_{u\in P_x} c_{u,\sigma}\right) 
\end{eqnarray}
and
\begin{eqnarray}
\left(\calP_{D}d_{x,-\sigma}^\dagger c_{x,\sigma}\right)
 \calP_{D}\zeta^\dagger
 &=&
 \calP_{D}(1-n_{x,-\sigma})
 d_{x,-\sigma}^\dagger c_{x,\sigma}\zeta^\dagger
\nonumber\\
&=&
 \calP_{D}(1-n_{x,-\sigma})
      d_{x,-\sigma}^\dagger \zeta^\dagger c_{x,\sigma}\nonumber\\
&&+ \sigma\alpha \calP_{D}(1-n_{x,-\sigma})
      d_{x,-\sigma}^\dagger 
      \sum_{u\in P_x} \rme^{2 \rmi Q\cdot u} a_{u,-\sigma}^\dagger
\nonumber\\
&=&
\label{eq:PdxbxPhi2}
 \calP_{D}(1-n_{x,-\sigma})
      \zeta^\dagger \left(\calP_{D} 
		     d_{x,-\sigma}^\dagger c_{x,\sigma}
		    \right).
\end{eqnarray}
Here we also used $\{c_{x,-\sigma},d_{x,-\sigma}^\dagger\}=0$ and
$(d_{x,-\sigma}^\dagger)^2=0$.
It follows {from} \eqref{eq:PdxbxPhi1} and \eqref{eq:PdxbxPhi2}
that ($\calP_D d_{x,-\sigma}^\dagger b_{x,\sigma})
 (\calP_D\zeta^\dagger)^{\frac{\Ne}{2}}\Phi_0 =0$, i.e., 
 $H_x\PhiGNe=0$
for all $x\in D$. Therefore we conclude that $H\PhiGNe = 0$.

The proof for the other statements in the theorem relies on the
following lemma, which will be proved later.
\begin{lemma}
 \label{lemma:1}
 Any zero-energy state $\Phi$ for $H$ 
 in the Hilbert space $\bigoplus_{\Ne=0}^{2|P|}\mathbf{H}_\Ne$ is written as
 \begin{equation}
  \Phi = \sum_{B\subset P}\phi(B)
    \calP_D\left(\prod_{u\in B}a_{u,\up}^\dagger\right) 
            \left(\prod_{u\in B}a_{u,\dn}^\dagger\right)\Phi_0, 
 \end{equation}
 where $\phi(B)$ are real coefficients.
 Furthermore coefficients satisfy  
 $|\phi(B)|=|\phi(B^\prime)|$ for any $B,B^\prime\subset P$ 
 such that $|B|=|B^\prime|$. 
\end{lemma}
It immediately follows {from} Lemma~\ref{lemma:1} 
that the ground state energy for 
odd $\Ne$ is always positive.
The remaining task is to prove the uniqueness.
Suppose that there are two zero-energy states for fixed even $\Ne$.
Then, an arbitrary linear combination of these states is also a
zero-energy state,
which should satisfy Lemma~\ref{lemma:1}.
However, we can make a suitable linear combination so that a
coefficient $\phi(B_0)$ for a subset $B_0$ will be vanishing, 
and this leads to the conclusion that 
all the other coefficients are also vanishing. 
This is contradicting with the assumption, and therefore the ground
state is unique.
This completes the proof of Theorem~\ref{theorem}. \qed 

Before proceeding to the proof of Lemma~\ref{lemma:1}, 
we prove the following lemma.
\begin{lemma}
\label{lemma:2}
For $A\subset D$, spin configuration $\bfs_A$ and 
$B_\up,B_\dn\subset P$, 
define
\begin{equation}
\label{eq:newbasis}
 \Phi(A,\bfs_A;B_\up,B_\dn)
 =\left(\prod_{x\in A}c_{x,\sigma_x}^\dagger\right) 
 \left(\prod_{u\in B_\up}a_{u,\up}^\dagger\right)
 \left(\prod_{u\in B_\dn}a_{u,\dn}^\dagger\right)\Phi_0.
\end{equation}
Then, the states $\calP_D\Phi(A,\bfs_A;B_\up,B_\dn)$ 
are linearly independent, and
the collection of these states with
 $|A|+|B_\up|+|B_\dn|=\Ne$
spans the Hilbert space $\mathbf{H}_\Ne$.
\end{lemma}
\textit{Proof of Lemma~\ref{lemma:2}}.     
Consider a linear combination
\begin{equation}
  \Phi = \sum_{A\subset D} \sum_{\bfs_A} \sum_{B_\up,B_\dn\subset P}
  \phi(A,\bfs_A;B_\up,B_\dn)\calP_D\Phi(A,\bfs_A;B_\up,B_\dn),
\end{equation} 
where $\sum_{\bfs_A}$ means the sum 
taken over all spin configurations.
We suppose that $\Phi=0$.
{From} the anticommutation relation
\eqref{eq:anticommutation1} 
it follows that 
the inner product
\begin{equation}
 \left(\Psi(A^\prime,\bfs_{A^\prime}^\prime;
  B_\up^\prime,B_\dn^\prime),
  ~\calP_D\Phi(A,\bfs_A;B_\up,B_\dn)\right)
\end{equation} 
is zero if both  $B_\up^\prime\subset B_\up$ 
and $B_\dn^\prime \subset B_\dn$ do not hold.
(See \eqref{eq:basis} for the definition of $\Psi(A,\bfs_A;B_\up,B_\dn)$.)
We furthermore have that    
\begin{equation}
  \left(\Psi(A^\prime,\bfs_{A^\prime}^\prime;B_\up,B_\dn),
  ~\calP_D\Phi(A,\bfs_A;B_\up,B_\dn)\right)
   =0
\end{equation}
for $A^\prime\ne A$ and that
\begin{equation}
  \left(\Psi(A,\bfs_{A}^\prime;B_\up,B_\dn),
  ~\calP_D\Phi(A,\bfs_A;B_\up,B_\dn)\right)
   =\chi[\mbox{$\sigma_x^\prime=\sigma_x$ for all $x\in A$}].
\end{equation}
Thus we obtain {from}  
$\left(\Psi(A,\bfs_A;P,P),
  ~\Phi\right)=0$
that $\phi(A,\bfs_A;P,P)=0$ for any $A\subset D$ 
and spin configuration $\bfs_A$. 
Then, examining 
\begin{equation}
\left(\Psi(A,\bfs_{A};P\bs\{u\},P),
  ~\Phi\right)=0
\end{equation}
for $u\in P$,
we find that $\phi(A,\bfs_{A};P\bs\{u\},P)$ = 0.
Similarly, examining
\begin{equation}
\left(\Psi(A,\bfs_{A};P\bs\bar{B}_\up,P\bs\bar{B}_\dn),
  ~\Phi\right)=0
\end{equation}
for $\bar{B_\up},\bar{B}_\dn\subset P$ 
with $|\bar{B}_\up|,|\bar{B}_\dn|=0,1,2,\dots|P|$ repeatedly,
we conclude that all the coefficients 
$\phi(A,\bfs_{A};{B}_\up,{B}_\dn)$ 
are vanishing.
This proves the first claim,
and the second claim is now trivial. \qed
\bigskip\\
\textit{Proof of Lemma~\ref{lemma:1}.} 
Suppose that $\Phi$ is an arbitrary zero-energy state for $H$ 
in $\bigoplus_{\Ne=0}^{2|P|}\mathbf{H}_\Ne$.
By using the basis states $\calP_D\Phi(A,\bfs_A;B_\up,B_\dn)$
with $|A|+|B_\up|+|B_\dn|\le 2|P|$, we represent $\Phi$ as 
\begin{equation}
 \Phi = \sum_{A\subset D} \sum_{\bfs_A} \sum_{B_\up,B_\dn\subset P}
  \phi(A,\bfs_A;B_\up,B_\dn)\calP_D\Phi(A,\bfs_A;B_\up,B_\dn).
\end{equation}
The sum is restricted to $|A|+|B_\up|+|B_\dn|\le 2|P|$,
but, here and in the following, 
we do not written down this restriction for notational simplicity.  

Since the local Hamiltonian $H_i$  
is the sum of two positive semidefinite operators, 
zero-energy state $\Phi$ for $H$ must be annihilated 
by these operators for any $i\in\La$.

We first examine the case $i=u\in P$.
In this case the condition $H_u\Phi=0$ is equivalent to
\begin{equation}
\label{eq:aucuPhi}
 \calP_D a_{u,-\sigma}^\dagger c_{u,\sigma}\calP_D\Phi = 0
\end{equation}
for $\sigma=\up,\dn$. 
For $\sigma=\up$ the left hand side of \eqref{eq:aucuPhi} becomes
\begin{eqnarray}
 &&\sum_{A\subset D} \sum_{\bfs_A} \sum_{B_\up,B_\dn\subset P}
  \sgn[u;B_\up,B_\dn]
  \chi[u\in B_\up]\chi[u\notin B_\dn]\nonumber\\
&&\hspace*{3cm}\times\phi(A,\bfs_A;B_\up,B_\dn)
  \calP_D\Phi(A,\bfs_A;B_\up\bs\{u\},B_\dn\cup\{u\}),
\label{eq:leftaucuPhi}
\end{eqnarray}
where $\sgn[u;B_\up,B_\dn]$ is a sign factor arising {from} 
exchanges of fermion operators. 
Since all the terms in~\eqref{eq:leftaucuPhi} 
are linearly independent,
we have the condition $\phi(A,\bfs_A;B_\up,B_\dn)=0$ 
for $B_\up,B_\dn$ such that 
$u\in B_\up$ and $u\notin B_\dn$.
A similar calculation for $\sigma=\dn$ yields
$\phi(A,\bfs_A;B_\up,B_\dn)=0$ for $B_\up,B_\dn$ such that 
$u\notin B_\up$ and $u\in B_\dn$.
Since these conditions must be satisfied for all $u\in P$, 
we obtain $\phi(A,\bfs_A;B_\up,B_\dn)=0$ for $B_\up\ne B_\dn$.

So far we have shown that a zero-energy state $\Phi$ 
can be expanded as
\begin{equation}
 \Phi = \sum_{A\subset D} \sum_{\bfs_A} \sum_{B\subset P}
  \phi(A,\bfs_A;B)\calP_D\Phi(A,\bfs_A;B,B),
\end{equation}
where $\phi(A,\bfs_A;B)=\phi(A,\bfs_A;B,B)$.
We further derive conditions on $\phi(A,\bfs_A;B)$ 
{from} $H_i\Phi = 0$ with $i=x\in D$, which is equivalent
to 
\begin{equation}
\label{eq:dxbxPhi}
 \calP_D d_{x,-\sigma}^\dagger b_{x,\sigma}\calP_D\Phi = 0
\end{equation}
for $\sigma=\up,\dn$.
The left-hand-side of \eqref{eq:dxbxPhi} can be expanded 
by using~\eqref{eq:newbasis}, so that we decompose this as
\begin{equation}
 \sum_{A^\prime\subset D;x\notin A^\prime}
 \sum_{\bfs_{A^\prime}^\prime}
 \sum_{B_\up^\prime,B_\dn^\prime\subset P} \cdots +
 \sum_{A^\prime\subset D;x\in A^\prime}
 \sum_{\bfs_{A^\prime}^\prime}
 \sum_{B_\up^\prime,B_\dn^\prime\subset P} \cdots 
\end{equation}
and write $\Phi_{x}$ for the first sum. 
Since the states in the first sum 
and those in the second sum
are linearly independent,
$\Phi_{x}=0$ must be satisfied.
 
To obtain $\Phi_{x}$ for $\sigma=\up$,
we operate $\calP_Dd_{x,\dn}^\dagger b_{x,\up}\calP_D$ 
on the basis state $\calP_D\Phi(A,\bfs_A;B,B)$.
Then, we have
\begin{equation}
(1-n_{x,\dn}) \calP_D d_{x,\dn}^\dagger c_{x,\up}\Phi(A,\bfs_A;B,B)
    +\beta \calP_D d_{x,\dn}^\dagger 
\left(\sum_{u\in P_x} c_{u,\up}\right)\Phi(A,\bfs_A;B,B).
\label{eq:dxbxPhi2}
\end{equation}
The second term in~\eqref{eq:dxbxPhi2} becomes
\begin{equation}
\beta 
   \sum_{u,u^\prime\in P_x}
   \sgn[u,u^\prime;B]
   \chi[u\in B]\chi[u^\prime \notin B] 
   \rme^{2\rmi Q\cdot u^\prime}
   \calP_D\Phi(A,\bfs_A;B\bs\{u\},B\cup\{u^\prime\}),
\end{equation}
where $\sgn[u,u^\prime;B]$ is a sign factor
arising {from} exchanges of fermion operators.
Thus, the second term contributes to $\Phi_{x}$ only when
$x\notin A$.
The first term in~\eqref{eq:dxbxPhi2} becomes
\begin{eqnarray}
\lefteqn{
(\chi[x\in A]\chi[\sigma_x=\up]+\chi[x\notin A])
\calP_D d_{x,\dn}^\dagger c_{x,\up}
\Phi(A,\bfs_A;B,B)
}
\nonumber\\
&&
-
(\chi[x\in A]\chi[\sigma_x=\up]+\chi[x\notin A])
\calP_D 
d_{x,\dn}^\dagger c_{x,\dn}^\dagger c_{x,\dn} c_{x,\up}
\Phi(A,\bfs_A;B,B).
\label{eq:dxbxPhi3}
\end{eqnarray}
Since site $x$ is always occupied by an electron in the second term in
\eqref{eq:dxbxPhi3},
this term never contributes to $\Phi_{x}$. 
The first term in \eqref{eq:dxbxPhi3} furthermore becomes
\begin{eqnarray}
&&
 \chi[x\in A]\chi[\sigma_x=\up]\sgn[x;A]\sum_{u^\prime\in P_x}
  \sgn[u^\prime;B]\chi[u^\prime\notin B]
\nonumber\\
&& \hspace*{5cm}
  \times\rme^{2\rmi Q\cdot u^\prime}
  \calP_D  \Phi(A\bs\{x\},\bfs_{A\bs\{x\}};B,B\cup\{u^\prime\})
\nonumber\\
&&+ \chi[x\in A]\chi[\sigma_x=\up]\calP_D 
 \left(\prod_{x\in A}c_{x,\sigma_x}^\dagger\right)
 d_{x,\dn}^\dagger c_{x,\up} 
 \left(\prod_{u\in B_\up}a_{u,\up}^\dagger\right) 
 \left(\prod_{u\in B_\dn}a_{u,\dn}^\dagger\right) 
 \nonumber\\
 &&+\chi[x\notin A]\alpha 
  \sum_{u,u^\prime\in P_x}
  \sgn[u,u^\prime;B]
  \chi[u\in B]\chi[u^\prime \notin B]
  \nonumber\\
 && \hspace*{5cm}
  \times 
  \rme^{2\rmi Q\cdot u^\prime}
  \calP_D\Phi(A,\bfs_A;B\bs\{u\},B\cup\{u^\prime\}),
\end{eqnarray}
where $\sgn[x;A]$ and $\sgn[u;B]$ are again sign factors.
Since site $x$ is always occupied by an electron 
in the second term in the above expression,
this term does not contribute to $\Phi_{x}$.
Therefore we finally obtain 
\begin{eqnarray}
\Phi_{x} &=& \sum_{A\subset D;x\in A}
                    \sum_{\bfs_A;\sigma_x=\up}
                    \sum_{B\subset P}
		    \sgn[x;A]
		   \sum_{u^\prime\in P_x}
		   \sgn[u^\prime;B]\chi[u^\prime\notin B]
\nonumber\\
 &&
  \hspace*{2cm}
  \times
  \rme^{2\rmi Q\cdot u^\prime}\phi(A,\bfs_A;B)
  \calP_D  \Phi(A\bs\{x\},\bfs_{A\bs\{x\}};B,B\cup\{u^\prime\})
  \nonumber\\ 
 &&+(\alpha+\beta)
  \sum_{A\subset D;x\notin A}
  \sum_{\bfs_A}
  \sum_{B\subset P} 
  \sum_{u,u^\prime\in P_x}
  \sgn[u,u^\prime;B]
  \chi[u\in B]\chi[u^\prime \notin B]
  \nonumber\\ 
 &&
  \hspace*{2cm}
  \times
  \rme^{2\rmi Q\cdot u^\prime}\phi(A,\bfs_A;B)
  \calP_D\Phi(A,\bfs_A;B\bs\{u\},B\cup\{u^\prime\}).
\label{eq:Phix}
\end{eqnarray}
It is noted that the terms in the first sum 
are linearly independent of those in the second sum. 

Choose a configuration $(A,\bfs_A;B)$ satisfying that
$A$ contains $x$, $\sigma_x$ in $\bfs_A$ is $\up$ and 
there exists $u^\prime \in P_x$ such that $u^\prime\not\in B$. 
Then, by checking the coefficient of the basis state 
$\calP_D\Phi(A\bs\{x\},\bfs_{A\bs\{x\}};B,B\cup\{u^\prime\})$ 
in \eqref{eq:Phix},
we obtain $\phi(A,\bfs_A;B)=0$ for such configuration.
Since this and a similar result for $\sigma=\dn$ must hold for any 
$x\in D$, we obtain that
\begin{equation}
\label{eq:condition1}
 \phi(A,\bfs_{A};B)=0 
\end{equation}
if $\left(\cup_{x\in A}P_x\right)$ is not a subset of $B$.

Next, choose a configuration satisfying that
$A$ does not contain $x$ and 
there exist $u,u^\prime \in P_x$ such that $u\in B$ 
and $u^\prime\notin B$.
For this kind of configuration,
by checking the coefficient of the basis state
$\calP_D\Phi(A,\bfs_{A};B\bs\{u\},B\cup\{u^\prime\})$ in \eqref{eq:Phix}, 
we have that
\begin{eqnarray}
 &&(\alpha+\beta)(
  \sgn[u,u^\prime;B]\rme^{2\rmi Q\cdot u^\prime}\phi(A,\bfs_A;B)
  \nonumber\\
 && \hspace*{2cm} +
  \sgn[u^\prime,u;B_{u\to u^\prime}]
  \rme^{2\rmi Q\cdot u}\phi(A,\bfs_A;B_{u\to u^\prime})
  )=0,
\end{eqnarray}
where $B_{u\to u^\prime} = \left(B\bs\{u\}\right)\cup \{u^\prime\}$.
Since $\sgn[\cdot],\rme^{2\rmi Q\cdot u}$ and 
$\rme^{2\rmi Q\cdot u^\prime}$ 
give only sign factors and
$\alpha+\beta$ is non-zero by definition, 
we obtain the (necessary) condition that,
for any $A$ not containing $x$ and any $\bfs_{A}$,
\begin{equation}
\label{eq:condition2}
 |\phi(A,\bfs_A;B)|=
  |\phi(A,\bfs_A;B_{u\to u^\prime})| 
\end{equation}
if there exist $u,u^\prime\in P_x$ such that $u\in B$ 
and $u^\prime\notin B$.

By using conditions \eqref{eq:condition1} and \eqref{eq:condition2},
we shall prove that $\phi(A,\bfs_A;B)=0$ for any $A\ne\emptyset$.
Suppose that $\phi(A,\bfs_A;B)\ne 0$ 
for some non-empty set $A\subset D$,
spin configuration $\bfs_A$ and $B\subset P$.
Here we say that $u$ and $u^\prime$ in $B$ are connected when
$D_u\cap D_{u^\prime}\ne\emptyset$ and decompose $B$ into
connected components as $B=B^1\cup\cdots B^m$.
We find {from} condition \eqref{eq:condition1} that
$A$ is contained in $\cup_{u\in B}D_u$, and
without loss of generality, we can assume that $\cup_{u\in B^1}D_u$ contains
at least one site in $A$. 
For connected subset $B^1$ 
we can find at least one pair of sites $u_0,u_1\in P$ 
and a site $x_1\in \cup_{u\in B^1}D_u$ 
such that 
$u_0\notin B^1,u_1\in B^1$ and $u_0,u_1\in P_{x_1}$.
When $m>1$ this is trivial, and 
when $m=1$ this follows {from} $\Ne = |A|+2|B|\le2|P|$. 
It is noted that $x_1$ is not in $A$ 
due to condition \eqref{eq:condition1}.
Then, since we can always find subsets 
$\{u_2,\dots,u_n\}\subset B^1$ and    
$\{x_2,\dots,x_n\}\subset \cup_{u\in B^1}D_u$ satisfying that
$u_{l}\notin B_{u_{l}\to u_0}^1$, $u_{l+1}\in B_{u_{l}\to u_0}^1$, 
$u_{l},u_{l+1}\in P_{x_{l+1}}$ and 
$x_{l} \notin A$ for $1\le l <n$,
and $x_n\in A$,
the repeated use of \eqref{eq:condition2} gives 
\begin{equation}
|\phi(A,\bfs_A;B)| = |\phi(A,\bfs_A;B_{u_{n-1}\to u_0})|.
\end{equation}
But, since $u_{n-1}\in P_{x_n}$ 
is not in $B_{u_{n-1}\to u_0}$,
the coefficient $\phi(A,\bfs_A;B_{u_{n-1}\to u_0})$ must be vanishing 
because of \eqref{eq:condition1}, which leads to a contradiction.
Therefore the claim is proved.

{From} the above results we conclude that $\Phi$ is expanded as
\begin{equation}
 \Phi = \sum_{B\subset P} \phi(B) 
  \calP_D\Phi(\emptyset,\bfs_{\emptyset};B,B),
\end{equation}
where $\phi(B)=\phi(\emptyset,\bfs_{\emptyset};B)$.
Then, using \eqref{eq:condition2} again, 
we find that $|\phi(B)|=|\phi(B^\prime)|$ whenever $|B|=|B^\prime|$,
which completes the proof of Lemma~\ref{lemma:1}. \qed       
\section*{Acknowledgement}
I would like to thank Kengo Tanaka for many discussions.

\end{document}